\magnification1200

\rightline{KCL-MTH-04-18}
\rightline{hep-th/0412336}

\vskip .5cm
\centerline
{\bf  Brane dynamics, central charges and $E_{11}$}
\vskip 1cm
\centerline{Peter West}
\centerline{Department of Mathematics}
\centerline{King's College, London WC2R 2LS, UK}

\leftline{\sl Abstract}
\vskip .2cm
\noindent 
We consider a  theory in which  supersymmetry is partially spontaneously
broken  and show that the dynamical fields in the same supersymmetric
multiplet as the Goldstino are Goldstone bosons whose corresponding 
generators  are central charges in the underlying supersymmetry algebra.
We illustrate how this works for four dimensional Born-Infeld theory  
and five brane of M theory. We conjecture, with supporting arguments, that
the dynamics of the  branes of M theory can be extended so as to possess
an $E_{11}$ symmetry.  
\vskip .5cm

\vfill
\eject
\medskip
{\bf {1. Introduction }}
\medskip
Given a theory with a symmetry algebra $G$
which is spontaneously broken to the sub-algebra $H$ then it is usually
the case that the low energy theory is a non-linear realisation of $G$
with local sub-algebra $H$ [1]. Hence, if one is  studying  a
phenomenon which is of this type then, by examining the low energy
behaviour,  one can discover  the symmetries of the  theory without
having to understand the underlying dynamics. Indeed, it was using this
idea   that some of the most important  symmetries of   particle physics
were first established. 
\par
One of the very first papers on
supersymmetry constructed  the non-linear realisation which results when
a four dimensional   $N=1$ supersymmetry theory  spontaneously breaks all
its of supersymmetry [7].  Superbranes can be thought of as a defects
in superspace in which supersymmetry is partially spontaneously broken and
as such should be described by a non-linear realisation. A selection
 of the substantial number of papers on this viewpoint, when half
 the supersymmetry is spontaneously broken,  are given in references
[1-6, 22]. Following these it has become 
well known that the fermion  field in superbrane dynamics is the
Goldstino for  broken supersymmetry and the transverse scalar fields are
the Goldstone boson for broken translations. In more recent years  branes
that include world volume gauge fields have played an important role. The
most important examples are   the D-branes in ten dimensions and the five
brane of M theory.   Such branes can be also  described as a non-linear
realisation  and it was shown [8] that the derivatives of the world
volume gauge fields and also the usual transverse scalars fields, are
 Goldstone bosons for  automorphism of the supersymmetry algebra.
However, this paper did not find a symmetry corresponding to the
world-volume gauge fields themselves.  
\par
In section two  we will consider a theory
in which supersymmetry is spontaneously broken and  present a simple
argument that shows that  the dynamical bosonic 
 fields in the same multiplet as the Goldstino are Goldstone
bosons whose corresponding   generators are  the  central
charges in the underlying supersymmetry algebra. This argument applies
to the world volume gauge fields of superbranes and we will show 
in section three  how this occurs in the general
non-linearly realised formalism and in particular in section four for 
the five
brane of M theory and the four
dimensional Born-Infeld theory.   Thus one finds that all the dynamical 
bosonic fields of branes  have a common origin in that they are the 
 Goldstone bosons corresponding to the central charges. 
\par
It has been proposed that M theory possesses an underlying $E_{11}$
symmetry; the low level fields in this non-linear realisation being
precisely those of the maximal supergravity theories [9]. The 
 latter are just the low energy description of the closed string
theory, however, it has been known since early on in the development of
string theory  that open string scattering leads to closed strings 
and so one might suspect that open strings, when suitably extended, 
should also possess an  $E_{11}$ symmetries. In particular, one might
hope that this would show up in the effective action for open
superstrings,  namely the Born-Infeld actions, or more generally in the
dynamics of the branes of M theory. 
\par
There is considerable evidence that all the usual brane charges 
are contained in a single fundamental representation of $E_{11}$ denoted
$l_1$ [11,14,16]. This  multiplet contains at its lowest levels the
space-time translations, and  the two and five form central charges
of the eleven dimensional supersymmetry algebra and then an infinite
number of more exotic objects. Regarding  branes as a result of  
spontaneously symmetry breaking it is a natural generalisation to think of
the dynamics of their bosonic fields as constructed from 
Goldstone bosons corresponding to generators which  belong to  the
$l_1$ representation. 
The result mentioned above for the five brane is an encouraging sign in
this direction as one finds that its dynamics contains the transverse
scalars and the world volume gauge field both of which occur as the first
two  fields in the $l_1$ representation and their interpretation as
Goldstone bosons for the central charges implies that they occur in the
non-linear realisation in a way that is a low level truncation of
an $E_{11}$ formulation. 
\par
However,  at first sight,  the situation for the two brane of M theory
appears less promising. However, in section five we sketch the low level
$E_{11}$ non-linear realisation based on the $l_1$ representation
appropriate to the two brane and recover the usual dynamics for the
bosonic fields. Finally, in section six we discuss some of the
implications of the conjecture advanced above.

%%%%%%%%%%%%%%%%%%%%%%%%%%%%%%%%%%%%%%%%%%%%%%%%%%%%%%%%%%%%%%%%%%%%%%%%
\medskip
{\bf {2. Central charges and Goldstone fields }}
\medskip
Let us consider a supersymmetric theory in which some, but not all, of the
supersymmetry is spontaneously broken. We denote  the preserved
supersymmetry generators by $Q_\alpha$ and the broken supersymmetry
generators by $Q_{\alpha^\prime}$. The indices $\alpha$ and
$\alpha^\prime$ may contain internal as  spinor indices. 
The Goldstino, that is the field corresponding to the
broken supersymmetry is denoted by $\bar \lambda^{\alpha^\prime}$ and
let
$A^\bullet$ be one of the dynamical bosonic fields in the same multiplet
as the Goldstino under the preserved  supersymmetry  $Q_\alpha$. Here
$\bullet$ labels any indices on the bosonic field. On dimensional and
Lorentz symmetry grounds it follows that the transformation of 
$A^\bullet$ under a preserved supersymmetry with parameter $\bar
\epsilon^{\alpha} $ is  of the form 
$$\delta  A^\bullet=\bar \epsilon^{\alpha}
(\gamma^\bullet C^{-1})_{\alpha\beta^\prime} 
\bar \lambda^{\beta^\prime}
\eqno(2.1)$$
Now by definition the Goldstino
transforms under the broken supersymmetry as 
 $\delta \bar  \lambda^{\alpha^\prime}=\eta^{\alpha^\prime}+\ldots $ 
where $+\ldots$ denotes field dependent terms. Consequently,  the
commutator of a broken and an unbroken supersymmetry on $A^\bullet$ takes
the form 
$$[\delta_\epsilon, \delta_\eta] A^\bullet=-\bar \epsilon^{\alpha}
(\gamma^\bullet C^{-1})_{\alpha \beta^\prime} 
\bar \eta^{\beta^\prime} +\ldots 
\eqno(2.2)$$
This implies that there is  a central charge $Z_\bullet$ in the underlying
supersymmetry algebra 
which occurs as 
$$
\{ Q_{\alpha}, Q_{\beta^\prime}\}=Z_\bullet(\gamma^\bullet
C^{-1})_{\alpha\beta^\prime}
\eqno(2.3)$$
The  action of $Z_\bullet$ on  $A^\bullet$ is to shift it by a 
constant and  consequently we can interpreted $A^\bullet$ as the 
Goldstone field  corresponding to the generator $Z_\bullet$. 
Thus we conclude that  a dynamical field in the same
supersymmetry multiplet as the Goldstino must be a  Goldstone boson
whose   corresponding     generator  is one of the central charges that
occurs in the anti-commutator of the preserved and broken
supersymmetries. 
\par
We can also identify the derivatives of the dynamical Goldstone fields. 
Let us assume that there exists a symmetry of the theory that 
rotates the Goldstino into the superspace coordinate $
\theta^\alpha$, namely 
$$
\delta \bar \lambda^{\alpha^\prime}= \theta^\beta
L_\beta {}^{\alpha^\prime}
\eqno(2.4)$$
Carrying out the commutator of a supersymmetry transformation and the
above rotation we find that 
$$
[\delta_\epsilon , \delta _L ] \bar \lambda^{\alpha^\prime}=
\bar \epsilon^\beta L_\beta  {}^{\alpha^\prime}+\ldots 
\eqno(2.5)$$
If we denote the  generator of the rotation by $R_{\alpha^\prime}{}^\beta
$ we must conclude that it  occurs in the underlying algebra as 
$$
[Q_\gamma ,R_{\alpha^\prime}{}^\beta ]=
\delta_\gamma ^\beta Q_{\alpha^\prime}
\eqno(2.5)$$
Hence it is one of the automorphism of the supersymmetry algebra studied
in [8]. 
\par
The Goldstino $\bar \lambda^{\alpha^\prime}$ and Goldstone
boson $A^\bullet$ belong to a supermultiplet whose first
component is 
$A^\bullet$. Denoting the superfield and its first component by the same
symbol equation (2.1) implies that $D_\alpha A^\bullet=
(\gamma^\bullet C^{-1})_{\alpha\beta^\prime} 
\bar \lambda^{\beta^\prime}$. 
As a consequence of
equation (2.4), the spinorial derivative of the Goldstino transforms as 
$$
\delta_L D_\alpha \lambda^{\beta^\prime}=   L_\alpha
{}^{\beta^\prime}+\ldots 
\eqno(2.6)$$
As such, the spinorial derivative of the Goldstino is the Goldstone boson
for the automorphism of the supersymmetry algebra with generator 
$R_{\alpha^\prime}{}^\beta $. However, the spinorial derivative, 
or equivalently the supersymmetry
variation,  of the
Goldstino contain the space-time derivatives of the dynamical Goldstone
fields and any auxiliary fields that my be present. As a consequence, we
conclude that the latter fields which do not occur with derivatives in
the supersymmetry variation, are also  Goldstone fields of some of the
automorphism of the supersymmetry algebra that mixes the preserved and
broken supercharges. 
\par
Let us also consider a rotation of the form 
$$\delta_{\hat L} A^\bullet
=\hat L ^\bullet {}_n X^n
\eqno(2.7)$$
where $X^n$ is the space-time coordinate of the theory and let $R_\bullet
{}^n$ be the  associated generator. The commutator of a space-time
translation with parameter $\zeta^n$ and the above rotation is 
$$
[\delta _\zeta , \delta_{\hat L} ]  A^\bullet=
\hat L ^\bullet {}_n \zeta^n+\ldots
\eqno(2.8)$$ 
This is a shift in $A^\bullet$ that is generated by the
previously identified generator $Z_\bullet$  and so we find that 
$$
[P_n, R_\bullet{}^m ]=\delta_n^m Z_\bullet
\eqno(2.9)$$
It follows that $\partial_n A^\bullet$ transforms as 
$\delta_{\hat L}\partial_n A^\bullet=\hat L ^\bullet {}_n+\ldots  $ and so
can be identified as the Goldstone boson whose generator is 
$R_\bullet{}^n$.
\par
This last result is consistent with our earlier considerations, since the 
super-Jacobi identity involving $Q_\alpha, Q_\beta$ and
$R_{\gamma^\prime}{}^\beta$ implies using equation (2.5) that 
$R_{\gamma^\prime}{}^\beta$ rotates  $Z_{\alpha\beta}\equiv \{Q_\alpha,
Q_\beta\}$ into $Z_{\alpha\gamma'}\equiv \{Q_\alpha,
Q_{\gamma'}\}$. Since $Z_{\alpha\beta}$ include the translation
generator 
$P_n$ certain of the $R_{\alpha^\prime}{}^\beta$ will rotate $P_n$ into
the central charges $Z_{\alpha\beta'}$ as in equation (2.9).
Expanding $R_{\alpha^\prime}{}^\beta$  in terms of Clifford algebra
elements we find a set of generators with vector indices which are
totally anti-symmetrised and as a result only   the totally
anti-symmetric part of
$R_\bullet{}^m$ can be identified with  part of
$R_{\alpha^\prime}{}^\beta$. 
In a given model only a sub-set of
the allowed possible central charges may be present  and 
 it can happen that some parts of
$R_{\alpha^\prime}{}^\beta$  do not induce any transformations on  
 $P_a$.  As such, the Goldstone
fields corresponding to these generators are  not 
space-time derivatives of the central charge Goldstone fields 
even though, as explained above, 
 they are the Goldstone bosons for
$D_\alpha\bar \lambda^{\beta^\prime}$. The obvious interpretation of
these Goldstone fields is that they  correspond to auxiliary fields. 
\par
In this section, we have not explicitly used the machinery of non-linear
realisations, however, in the next section we will give the theory of
non-linear realisations applied to the current context and we will see
how the results found in this section emerge from the general formalism. 
\par
This interpretation of auxiliary fields as part of the automorphism
algebra may provide a new way of finding auxiliary fields.  
%%%%%%%%%%%%%%%%%%%%%%%%%%%%%%%%%%%%%%%%%%%%%%%%%%%%%%%%%%%%%%%%%%%%%%%%
\medskip
{\bf {3. The general formalism }}
\medskip

We first recall the description of branes as non-linear realisations as
given in reference [8]. We consider the supersymmetry algebra,  
which  obeys
the relations
$$\{Q_{\underline \alpha} ,Q_{\underline \beta} \}
=Z_{\underline \alpha\underline \beta},\qquad
[Q_{\underline \gamma},Z_{\underline \alpha\underline \beta}] = 0,
\qquad
[Z_{\underline \alpha\underline \beta},
Z_{\underline \gamma\underline \delta}]=0,
\eqno(3.1)$$
and denoted it by    
$\underline K$, 
We also consider    the
automorphism algebra 
$\underline H$ of $\underline K$ which obeys the relations 
$$
[Q_{\underline \alpha}, R_{\underline \gamma}{}^{\underline \delta}]=
 \delta_{\underline \alpha}{}^{\underline \delta} Q_{\underline
\gamma},\  [Z_{\underline \alpha \underline \beta}, R_{\underline
\gamma}{}^{\underline \delta}]
= \delta_{\underline \alpha}{}^{\underline \delta}
Z_{\underline \gamma
\underline \beta} + \delta_{\underline \beta}{}^{\underline \delta}
Z_{\underline \alpha \underline \gamma}.
\eqno(3.2)$$
The generators of $\underline K$ and  $\underline H$ together form  the
algebra 
$\underline G$ from which the non-linear realisation is constructed. The
central charge generators
$Z_{\underline
\alpha\underline
\beta}$  include the space-time momentum
generators  $P_{\underline a}$.   It is easy to verify that such an
algebra obeys the generalized super Jacobi identities.  
\par
Expanding $Z_{\underline
\alpha\underline \beta}$ out in terms of the
enveloping algebra of the 
relevant Clifford algebra we
find that it contains a set of
  generators which are totally anti-symmetric in their vector
indices:
$$
Z_{\underline \gamma}{}^{\underline \delta}= \sum_p \sum_{\underline
n_1\ldots
\underline n_p} (\gamma^{\underline n_1\ldots
\underline n_p}C^{-1})_{\underline \gamma}{}^{\underline \delta }
Z_{\underline n_1\ldots \underline n_p}.
\eqno(3.3)$$
Similarly, 
we may expand $R_{\underline \gamma}{}^{\underline \delta}$, namely 
$$
R_{\underline \gamma}{}^{\underline \delta}= \sum_p \sum_{\underline
n_1\ldots
\underline n_p} (\gamma^{\underline n_1\ldots
\underline n_p})_{\underline \gamma}{}^{\underline \delta }
R_{\underline n_1\ldots
\underline n_p}.
\eqno(3.4)$$

If all possible central charges are allowed the generators
$Z_{\underline
\alpha\underline \beta}$ form the most general symmetric matrix and 
the automorphism group is $GL(c_d)$ where $c_d$ is
the number  of supercharges. However, it is often the case that we
require only a sub-set of all the possible  central charges in which case
the  automorphism group is reduced. Indeed, 
if   the only central charge is the momentum then
the right-hand side of the anti-commutator of two
supercharges is
$\gamma^{\underline a}P_{\underline a}$, then  the most
general automorphism group is  by definition the spin group. This natural
enlargement of the Lorentz algebra when more central charges are 
present acts as a brane rotating symmetry [8,15].  
\par
Let us consider the action of one of the  automorphisms $R^\bullet$ 
that occurs in equation (3.4) where $\bullet$ encodes the indices.
Equation (3.2) then takes the form 
$$[Z_{\underline \alpha \underline \beta}, R^\bullet ]
=(\gamma^\bullet )_{\underline \alpha}{}^{\underline \delta} Z_{\underline
\delta \underline \beta}+
(\gamma^\bullet )_{\underline \beta}{}^{\underline \delta}
Z_{\underline \alpha\underline\delta }
\eqno(3.5)$$
We may rewrite this equation as 
$$
[Z_{\underline \alpha \underline \beta}, R^\bullet ]=
(\gamma^\bullet ZC-\eta\epsilon ZC\gamma^\bullet)_{\underline \alpha}
{}^{\underline \delta} (C^{-1})_{\underline \delta \underline \beta}
\eqno(3.6)$$
where the charge conjugation matrix $C$ satisfies $C^T=-\epsilon C$ and 
$(\gamma^\bullet C^{-1})^T=\eta \gamma^\bullet C^{-1}$. Given a set of 
central charges $Z_{\underline \alpha \underline \beta}$ we may use the
last equation to find what is the maximal  automorphism algebra. We note
that it depends on the transposition  properties of $\gamma^\bullet$ and 
$C$ both of which are dimension dependent.
\par
We divide the  generators in $\underline K$ into
$Q_{\underline \alpha}= (Q_\alpha ,Q_{\alpha'})$
and
$Z_{\underline \alpha \underline \beta}=( Z_{ \alpha \beta}, Z_{
\alpha \beta'}, Z_{ \alpha' \beta'})$ and the generators of
$\underline H$ as $R_{\underline \gamma}{}^{\underline \beta}
=(R_{ \gamma}{}^{ \beta},R_{ \gamma'}{}^{ \beta},
R_{ \gamma}{}^{ \beta'},R_{ \gamma'}{}^{ \beta'})$. 
In this  decomposition $Q_\alpha$ and $Q_{\alpha^\prime}$ are the
preserved  and broken supersymmetry generators respectively. The
decomposition of the spinor indices corresponding to the breaking of  the
underlying spin algebra.  
The generators   $Q_\alpha$, $Z_{\alpha\beta}$ and a suitable 
sub-algebra $H$ of $\underline H$ are preserved and we denote the
algebra of these generators by $G$. 
\par
We consider the non-linear realisation of n $\underline G$ with local
sub-algebra $H$ and so consider the group element 
$$g= e^{X^{\alpha \beta}Z_{\alpha \beta}+\theta ^\alpha Q_\alpha}
e^{X^{\alpha \beta'}Z_{\alpha \beta'}+X^{\alpha' \beta'}Z_{\alpha'
\beta'}+\Theta ^{\alpha'} Q_{\alpha'}}e^{\phi \cdot R }
$$
$$\equiv g_p e^{X^{\alpha \beta'}Z_{\alpha \beta'}+X^{\alpha'
\beta'}Z_{\alpha'
\beta'}+\Theta ^{\alpha'} Q_{\alpha'}}e^{\phi \cdot R }
\eqno(3.7)$$
where $\phi\cdot R=\phi_{\underline \delta}{}^{\underline\gamma}
R_{\underline\gamma}{}^{\underline\delta}$ is a sum that includes all the
generators in
$\underline H$, except for those in $H$.  
Although the group element contains $Z_{\alpha \beta}$ and $Q_\alpha$
these do correspond to preserved symmetries and their role is to
introduce superspace into the theory. In particular , $X^{\alpha
\beta'}$,  $X^{\alpha' \beta'}$, $\Theta ^{\alpha'}$ and $\phi$
depend on this superspace.  
\par
To illustrate the results of the previous section,  it will suffice to
consider the linearised approximation in which we  keep terms
only to first order in the dynamical fields. Hence, $X^{\alpha \beta}$ and
$\theta^\alpha $ are of order zero and all other fields are of order
one.   To this order the Cartan forms are given by 
$$g^{-1}d g= dz^\pi (E_\pi {}^a P_a + E_\pi {}^\alpha Q_\alpha)+
E^N\nabla_N \Theta^{\alpha^\prime} Q_{\alpha^\prime}
$$
$$
+ E^N\nabla_N X^{\alpha \beta'}Z_{\alpha \beta'}
+E^N\nabla_N X^{\alpha' \beta'}Z_{\alpha'
\beta'}+
E^N\nabla_N \phi\cdot R 
\eqno(3.8)$$
where
$$ g^{-1} _p d g_p\equiv dz^\pi (E_\pi {}^a P_a + E_\pi {}^\alpha
Q_\alpha) \equiv E^a P_a + E^\alpha Q_\alpha=d\theta^\alpha Q_\alpha 
$$
$$+
(dX^a-{1\over 2}d\theta^{\alpha} (\gamma^a C^{-1}\theta)_\alpha)P_a,
\eqno(3.9)$$
and 
$$
\nabla_{\gamma}X^{\alpha\beta '}= D_{\gamma}X^{\alpha\beta '}
-\delta_{\gamma}^{\alpha}\Theta^{\beta '}
\eqno(3.10)$$
$$\nabla_{\alpha}\Theta^{\beta'}=D_{\alpha}\Theta^{\beta'}+
\phi_{\alpha}{}^{\beta'}
\eqno(3.11)$$
$$\nabla_{\gamma\delta} X^{\alpha\beta '}= \partial_{\gamma\delta}
X^{\alpha\beta '} + {1\over 2}(\delta_{\gamma}^{\alpha}
\phi_{\delta}{}^{\beta'} + \delta_{\delta}^{\alpha}
\phi_{\gamma}{}^{\beta'})
\eqno(3.12)$$
We also find that $
\nabla_{\gamma\delta}\Theta^{\alpha'}
=\partial_{\gamma\delta}\Theta^{\alpha'}$ and  $\nabla_N \phi_{\underline
\alpha}{}^{\underline \beta} =D_N \phi_{\underline \alpha}{}^{\underline
\beta}$, although these two quantities will play
little further role in what follows. 
In the above we have  denoted the coordinates of superspace by  
$z^\pi=(X^a, \theta^\alpha )$,  the index range of $N=(a, \alpha)$ and 
the world indices, such as $\pi$, by the same range since we are working 
with the linearised approximation.  
In deriving these results we have defined the covariant derivative in 
superspace by 
$$d f=d z^\pi \partial_\pi f=E^N D_N f
\eqno(3.13)$$ 
for any function $f$ of superspace and where $E^N=dz^\pi E_\pi {}^N$. 
It is a consequence of the linearized analysis that
$\phi_{\gamma}{}^{\beta'}$  is the only part of
$\phi_{\underline \gamma}{}^{\underline \beta}$ that enters any of
these expressions.
\par
The dynamics is given by setting 
$$
\nabla_{\gamma}X^{\alpha\beta '}=0=\nabla_{\alpha}\Theta^{\beta'}\ \ {\rm
or\  equivalently }\ \ D_{\gamma}X^{\alpha\beta '}
=\delta_{\gamma}^{\alpha}\Theta^{\beta '}, \ 
D_{\alpha}\Theta^{\beta'}=-
\phi_{\alpha}{}^{\beta'}
\eqno(3.14)$$
The Cartan forms transform under $H$ and given our choice of automorphism
algebra  the above set of constraints is invariant. 
\par
The constraints of equation (3.14) make contact with the general
discussion of the previous section. The first of these equations implies
that  the supersymmetry variation of  
$X^{\alpha\beta '}$, denoted
$A^\bullet$ in the previous section, has the form of  equation (2.1). 
This equation also is often sufficient to determine the dynamics of the
brane.   The second of
these equations tells us that the spinorial derivative of the Goldstino
is part of the automorphism algebra and so belongs to ${\underline H/
H}$.  By taking the spinorial covariant derivatives of the first equation 
we find an alternative expression for the spinoral covariant derivative
of the Goldstino which involves  the space-time derivatives of
$X^{\alpha\beta '}$ and as such we conclude that the latter are part of
the automorphism algebra [8]. 
This can be viewed as an example of the inverse Higgs mechanism [23]. 
Although in reference [8] a general form for $X^{\alpha\beta '}$ was
allowed   in the  general formalism  it was then assumed  that  these
fields
 contained only the transverse fields $X^{a'}$ or, equivalently that  the
only active central charges were the space-time momenta.  However, in this
paper it was  also suggested  [8]  that one should take a more general
form for $Z_{\alpha\beta '}$, or 
$X^{\alpha\beta '}$, and it is this suggestion that 
 is implemented here. 
In the examples given below we will  find that for some theories these
extra generators  play an important role in that their corresponding
fields are  the world volume gauge fields that occur in some branes.  

%%%%%%%%%%%%%%%%%%%%%%%%%%%%%%%%%%%%%%%%%%%%%%%%%%%%%%%%%%%%%%%%%%%%%%%%
\medskip
{\bf 4. Examples }
\medskip
Certain aspects of the general scheme set out above have occurred in a
number of papers in the literature. In particular, it is well known
following the two- and four-dimensional examples worked out reference
[2-6] that the transverse scalar fields that occur in  brane actions are
the Goldstone bosons for broken translations. It is obvious, but not
always stressed  that  these broken translations must occur in the
supersymmetry algebra of superbranes as central charges.  A more recent
example of this phenomenon is the  construction of the $N=1, D=4$
supermembrane [5]. 
\par
The non-linear realisation that leads to a chiral superfield in four
dimensions, whose components we  denote by $({\cal
A};\chi;{\cal F})$,     was worked out in [4,3]. In these papers
the authors considered  a four dimensional
$N=2$ supersymmetry algebra with one complex scalar central charge $Z$ 
which   could  be viewed as a six dimensional supersymmetry
algebra, the complex central charge playing the role of the additional
momenta.  This  supersymmetry algebra possessed the  preserved generators
$P_a$, $Q_A, Q_{\dot A}$ and $J_{ab}$ and broken  generators $S_A, Q_{\dot
A}$, $Z$. Its automorphism algebra  was SO(1,5)$\otimes$ SU(2)  broken  to
SO(1,3)$\otimes$ SO(2)$\otimes$U(1). They found that   the Goldstino is
the fermion of the chiral multiplet $\chi_A, \chi_{\dot A}$, the Goldstone
boson corresponding to the central charge is the complex scalar ${\cal
A}$ and the Goldstone boson for the broken SU(2) part of the
automorphism algebra are the complex  auxiliary field
${\cal F}$.  Furthermore,  they noted that the remaining broken
automorphisms were the space-time derivatives of the ${\cal A}$. 
This particular example possesses many features of the general scheme set
about above, with the exception, like all previous work, that  it involves
only scalar central charges.  
\par
Below we will give three examples  that illustrate the general scheme. 
We first give the M2 brane for comparison and then give the M five brane
and finally the four dimensional Born-Infeld theory. Both the latter
cases possess  gauge fields which are the Goldstone bosons corresponding
to  central charges. 
\medskip
{\bf 4.1 The M2 brane}
\medskip
We briefly recall how the M2 brane works. 
In  this case the underlying algebra is 
$$\{Q_{\underline \alpha} ,Q_{\underline \beta} \}
=(\gamma^{\underline a}C^{-1})_{\underline \alpha\underline
\beta}P_{\underline a}
\eqno(4.1.1)$$
Hence the only central charges are the translations and correspondingly
we take 
$ X^{\underline
\alpha\underline
\beta }=(C
\gamma^{\underline a})^{\underline \alpha\underline \beta }X_{\underline
a}$. The maximal automorphism algebra $\underline H$  is just the Lorentz algebra SO(1,10), or  strictly
speaking the spin algebra spin  (1,10) for which 
$R_{\underline \gamma}{}^{\underline \delta}=  \sum_{\underline
a\underline b} (\gamma^{\underline a \underline b})_{\underline
\gamma}{}^{\underline \delta } J_{\underline a\underline b}$. 
Hence,   the only non-zero $\phi_{\underline \gamma}{}^{\underline
\delta}$ are the    $\phi _\alpha{}^{\beta '}=\sum_{ab'}
(\gamma ^{ab'})_{\alpha}{}^{ \beta'}\phi_{ab'}$. The preserved
sub-algebra of spin(1,10) is $H=$ spin(1,2)$\otimes$ spin (8). 
 \par
The  constraints of equation (3.14) transform covariantly  under
spin(1,2)$\otimes$ spin (8) and so adopting these gives a set of
equations which is invariant under the full non-linearly realised
algebra.  The first constraint implies 
that 
$$ D_\gamma X^{a'}={1\over 16}(\gamma^{a'}C^{-1})_{\gamma\delta'
}\Theta^{\delta'} ,
\eqno(4.1.2)$$
which is indeed implies the correct equations of motion  for the
linearised dynamics.   While the second equation implies that 
$$
D_\alpha \Theta^{\beta^\prime}=-(\gamma
^{ab'})_{\alpha}{}^{\beta'}\phi_{ab'}. 
\eqno(4.1.3)$$
Applying a spinorial covariant derivative to equation (4.1.2) 
and using equation (4.1.3) we find that 
$\partial_m X^{n'}\sim-\phi_{m}{}^{n'}$ confirming that  the 
space-time derivatives of
$X^{a'}$ are the Goldstone boson
$\phi_{ab'}$ corresponding to the automorphism $J_{ab'}$. In other words
the space-time derivatives of $X^{a'}$, i.e. $\partial _bX^{a'}$ belong
to the coset spin(1,10)/spin(1,2)$\otimes$spin(8). We also find that the
constraints of equations (4.1.2) and (4.1.3) imply that $\nabla_m
X^{a'}=0$
\medskip
{\bf 4.2 The M5 brane}
\medskip
The underlying supersymmetry algebra ${\underline K}$ for the five brane
is 
$$\{Q_{\underline \alpha} ,Q_{\underline \beta} \}
=(\gamma^{\underline a}C^{-1})_{\underline \alpha\underline
\beta}P_{\underline a}+
(\gamma^{\underline a\underline b}C^{-1})_{\underline \alpha\underline
\beta}Z_{\underline a\underline b}
+
(\gamma^{\underline a_1\ldots \underline a_5}C^{-1})_{\underline
\alpha\underline
\beta}Z_{\underline  a_1\ldots \underline a_5}
\eqno(4.2.1)$$
Although we consider $P_{\underline a}$ for $\underline a=0,\ldots ,10$
we will take the other central charges to carry indices only over the
range $\underline a=0,\ldots ,5$.  We must decompose the eleven 
dimensional Clifford
algebra into one that keeps manifest the Clifford algebra appropriate
to the five brane i.e spin(1,10) into spin(1,5)$\otimes$ spin(5). This
results in a corresponding decomposition of the spinor index $\underline
\alpha= (\alpha, \alpha')$ and then $\chi_{\alpha}\to \chi_{\alpha i}$
and $\chi_{\alpha'}\to \chi^{\alpha}_{ i}$. For the fivebrane
$\alpha=1,\dots ,4$ are the Weyl projected spinor indices of
Spin(1,5) and $i=1,\dots ,4$ are the indices of the internal group
$Usp(4)=Spin(5)$. 
\par
Applying this decomposition to the supersymmetry algebra we find that
the  preserved supercharges $Q_{\alpha i}$ obey the anti-commutator  
$$
[Q_{\alpha i}, Q_{\beta j} ]=\eta_{ij}(\gamma^a)_{\alpha \beta}P_a +
\eta_{ij}(\gamma^{a_1\ldots a_5}) Z_{a_1\ldots a_5}=
\eta_{ij}(\gamma^a)_{\alpha \beta}\hat P_a 
\eqno(4.2.2)$$
where $\hat P_a=P_a-\epsilon _a{}^{ b_1\ldots b_5}Z_{b_1\ldots b_5}$.
Although we have a five form in the preserved supersymmetry
algebra it can be absorbed  to leave just a
usual translation. However, the five form will reappear in the
supersymmetry algebra for two broken supersymmetry generators. 
$$
[Q^{\alpha}{}_ i, Q^{\beta}{}_ j ]=\eta_{ij}(\bar \gamma^a)^{\alpha
\beta}(P_a +\epsilon _a{}^{ b_1\ldots
b_5}Z_{b_1\ldots b_5} )
\eqno(4.2.3)$$
For the anti-commutator of a broken and an unbroken supersymmetry we find
that 
$$
[Q_{\alpha i}, Q^{\beta}{}_ j ]=(\gamma^{a'})_{ij}\delta_\alpha
^\beta P_{a'} +
\eta_{ij}(\gamma^{a_1 a_2})_\alpha
{}^\beta Z_{a_1 a_2} .
\eqno(4.2.4)$$
\par
The general decomposition $X^{\underline \alpha \underline \beta}=X^\bullet
(C\gamma_\bullet)^{\underline \alpha \underline \beta}$ takes on a
restricted form corresponding to the form of the anti-commutators in
equation (4.2.2), (4.2.3) and (4.2.4). In particular  in the world
volume we take 
$X^{ \alpha \beta}\equiv  X^{ \alpha i\beta j}=\hat X^a \eta^{ij} (\bar
\gamma_a)^{\alpha \beta}$ and so   we  have in effect just the usual
coordinates of space-time.  Corresponding to equation (4.2.4), we  
take 
$$
X^{ \alpha \beta'}\equiv  X^{ \alpha i}{}_\beta {}^j=-X^{a'}
(\gamma_{a'})^{i j} \delta^\alpha_\beta+ \eta^{ij}
(\bar \gamma^{ab})^{\alpha}{}_\beta B_{ab}
\eqno(4.2.5)$$
We will recognise $X^{a'}$ as the transverse scalars of the five brane
and $B_{ab}$ as its world volume gauge field. 
\par
Taking into account the restricted range of the indices on the central
charges, the automorphism algebra of the supersymmetry algebra of 
equation (4.2.1) is generated by 
$$
R_{\underline \alpha}{}^{\underline \beta}=
J_{\underline a \underline b}(\gamma^{\underline a \underline
b})_{\underline \alpha}{}^{\underline
\beta}+
R_{a_1a_2a_3}(\gamma^{a_1a_2a_3})_{\underline \alpha}{}^{\underline
\beta}
\eqno(4.2.6)$$
where $R_{a_1a_2a_3}$ is anti-self dual. In particular, we have 
$R_{\alpha'}{}^{\beta}==J_a{}^{b'}
(\gamma_{b'})_i{}^j(\bar \gamma^{a})^{\alpha\beta}+
R_{a_1a_2a_3}\delta_i^j(\bar \gamma^{a_1a_2a_3})^{\alpha\beta}
$. 
Indeed,
using equation (3.6) one can verify that once one introduces the two form
central charge one must also include the five form central charge. The
corresponding Goldstone fields required at lowest order 
 have the form 
$$
\phi_{ \alpha}{}^{ \beta'}\equiv 
\phi_{\alpha i}{}_{\beta}{}^j=-\phi_a{}^{b'}
(\gamma_{b'})_i{}^j(\gamma^{a})_{\alpha\beta}+
\phi_{a_1a_2a_3}\delta_i^j(\gamma^{a_1a_2a_3})_{\alpha\beta}
\eqno(4.2.7)$$
where $\phi_{a_1a_2a_3}$ is self-dual. 
\par
Using equation (4.2.5), equation
(3.10) implies that  the  equation 
$$
D_{\gamma k} X^{a'}= {1\over 16} (\gamma^{a'})_k{}^{j}
\Theta _{j\gamma }
\eqno(4.2.8)$$
and also 
$$
D_{\gamma k}B_{ab}={1\over 16}(\gamma_{ab})^\beta{}_\gamma
\Theta_{k\beta} ,
\eqno(4.2.9)$$
both of which are consistent with equation (2.1). 
Using equation (4.2.7), the condition of equation (3.11) then implies 
$$
D_{\alpha i} \Theta ^\beta{}_j
=-\phi_a{}^{b'}
(\gamma_{b'})_i{}^j(\gamma^{a})_{\alpha\beta}+
\phi_{a_1a_2a_3}\delta_i^j(\gamma^{a_1a_2a_3})_{\alpha\beta}
\eqno(4.2.10)$$
Equation (4.2.8) ensures  the correct dynamics for the five brane 
[21,25] and  in particular implies that 
$$ 
D_{\alpha i}\Theta _\beta {}^{ j}\sim
-(\rlap/ \partial)_{\alpha \beta}(\gamma_{n'})_i{}^{j} X^{n'}
+\delta ^j_i (\gamma_{n_1n_2n_3})_{\alpha \beta} h^{n_1n_2n_3}
\eqno(4.2.11)$$
where $h_{n_1n_2n_3}$ is the self-dual gauge field strength of the
fivebrane. Hence , we find that the space-time derivatives of the scalar
and  second rank gauge fields are identified with the automorphism group 
[8]. The underlying consistency of equations (4.2.8) and  (4.2.9) 
are ensured  by existence of the (2,0) supermultiplet  which they
specify.  We note that as $B_{ab}$ has a gauge symmetry, the supersymmetry
algebra will close only if one includes such transformations. 
\medskip 
{\bf 4.3 The four-dimensional Born-Infeld Theory}
\medskip
The underlying supersymmetry algebra ${\underline K}$ for this model when
written in two component notation is 
$$ \{Q_A, Q_B\}=0=\{S_A, S_B\},\  \{Q_A, Q_{\dot
B}\}=-2i(\sigma^c)_{A\dot B} P_c= 
\{S_A, S_{\dot B}\},
\eqno(4.3.1)$$
and 
$$
\{Q_A, S_{\dot B}\}=-2i(\sigma^c)_{A\dot B} Z_c. 
\eqno(4.3.2)$$
The automorphism algebra contains the usual Lorentz rotations 
$$ [Q_A, J_{ab}]= {1\over 2}(\sigma_{ab})_{A}{}^{ B} Q_B,\ 
[S_A, J_{ab}]= {1\over 2}(\sigma_{ab})_{A}{}^{ B} S_B
\eqno(4.3.3)$$
and well as automorphisms with generators $R_{ab}$ and $R$ which obey the
commutators 
$$ [Q_A, R_{ab}]= {1\over 2}(\sigma_{ab})_{A}{}^{ B} S_B,\ 
[S_A, R_{ab}]= {1\over 2}(\sigma_{ab})_{A}{}^{ B} Q_B,\ 
[Q_A, R]=S_A,\ [S_A, R]=Q_A
\eqno(4.3.4)$$
The super Jacobi identities determine the remaining commutators
to be 
$$
[R_{ab}, R_{cd}]= J_{ad}\eta_{bc}+\ldots ,\ 
[R_{ab}, J_{cd}]= R_{ad}\eta_{bc}+\ldots ,\ [R_{ab}, R]=0, \ 
[J_{ab}, R]=0,
\eqno(4.3.5)$$
and 
$$[P_c, R_{ab}]=-\eta_{ac}Z_b+\eta_{bc}Z_a,\ 
[Z_c, R_{ab}]=-\eta_{ac}P_b+\eta_{bc}P_a ,\ 
[P_c, R]=2Z_c,\ [Z_c, R_{ab}]=2P_c
\eqno(4.3.6)$$
where in these equations $+\ldots$ means one must add the corresponding
terms required by anti-symmetry. We note that this algebra has an
additional central charge
$Z_a$ compared to the usually used $N=2$ supersymmetry algebra and this
allows the presence of the additional automorphism $R_{ab}$. The
generators $R$ is just part of the usual SU(2)$\otimes$U(1) $R$ symmetry 
of $N=2$ supersymmetry. 
\par
Leaving aside the generator $R$, the above  algebra is just two copies of
the standard
$N=1$ supersymmetry algebra. If we define 
$$ \psi^{\pm}_A=Q_A\pm S_A,\ P_a^\pm=P_a\pm Z_a,\
J_{ab}^\pm={1\over 2}(J_{ab}\pm R_{ab})
\eqno(4.3.7)$$
they  obey
$$\{  \psi^{\pm}_A,  \psi^{\pm}_{\dot B}\}=-2i(\sigma^c)_{A\dot
B}P^\pm_c,\ 
\{  \psi^{\pm}_A,  \psi^{\pm}_{ B}\}=0,
$$
$$ [ \psi^{\pm}_A,
J_{ab}^\pm ]= {1\over 2}(\sigma_{ab})_{A}{}^{ B} \psi^\pm_B,\ 
[P^\pm_a, J_{bc}^\pm ]= - \eta_{ab}P^\pm_c+ \eta_{ac} P^\pm_ b
\eqno(4.3.8)$$
The two copies of the $N=1$ supersymmetry algebras (anti-)commute with
each other. The commutator with $R$ is $[\psi^\pm , R]=\pm
\psi^\pm$.  Two separate algebras
are mixed together in the dynamics as the preserved space-time
translations,   supercharges
and  Lorentz algebra are a superposition of generators from the two
separate algebras. 
\par
Starting from the algebra of equation (4.3.1) and (4.3.2), but with $Z_a$ 
absent and no automorphism generators as given in equation (4.3.4), except
the Lorentz algebra of  equation (4.3.3),  the full four dimensional Born
Infeld theory was constructed [3] as a non-linear  realisation. These
authors also suggested that  the auxiliary field present in this model was
related to the part of the usual R symmetry algebra.  
\par
We now consider the non-linear realisation of the algebra of equations 
(4.3.1), (4.3.2) and (4.3.3) with the local sub-algebra being just the
Lorentz algebra with generators $J_{ab}$. In the notation of the
beginning of this section the preserved   supercharges are
$Q_\alpha=(Q_A, Q^{\dot A})$ while the broken supercharges are
$Q_{\alpha'}=(S_A, S^{\dot A})$. The corresponding group element
takes the from 
$$
g=exp(x^a P_a+\theta^A Q_A+ \theta^{\dot A} Q_{\dot A})
exp(W^A S_A +W^{\dot A} S_{\dot A})exp (A^a Z_a)
exp(\phi^{ab} R_{ab} +\phi R) 
\eqno(4.3.9)$$
The Cartan forms of equation (3.10) of the linearised theory become in
this case 
$$\nabla_A A^a= D_A A^a +2i (\sigma^a)_{A\dot B} W^{\dot B},\ 
\nabla_{\dot A} A^a= D_{\dot A} A^a +2i (\bar \sigma^a)_{\dot A B}
W^{ B},\ 
\eqno(4.3.10)$$
While those of equation (3.11) become 
$$
\nabla_B W^A=D_B W^A+ {1\over 2} (\sigma^{ab})_B{}^A 
\phi _{ab}+
\delta_A^B \phi,\ \nabla_{\dot B} W^A=D_{\dot B} W^A,
$$
$$
\nabla_{\dot B} W^{\dot A}=D_{\dot B} W^{\dot A}- {1\over 2}
(\sigma^{ab})_{\dot B}{}^{\dot A} \phi _{ab}+
\delta_{\dot A}^{\dot B} \phi,\ \nabla_{ B} W^{\dot A}=D_{ B} W^{\dot
A},
\eqno(4.3.11)$$
The Grassmann even  constraints of equation (3.14) then imply that 
$$
D_{\dot B} W^A=0= D_B W^{\dot A},\ 
{\rm and }\  D_B W^B=D_{\dot B} W^{\dot B}
\eqno(4.3.12)$$
where we have taken into account of the reality of $\phi$. We note that in
this case the Grassmann even constraints have not only determined the
automorphism Goldstone bosons of the theory, but they have also provided
the conditions that specify the gauge covariant theory. 
Equations (4.3.11) and  (3.4.12)  are the correct  linearised
superspace constraints of the Born-Infeld theory off shell. In particular
they imply  that the field strength and auxiliary field of  the theory
are  identified with the Goldstone bosons for the automorphism $\phi_{ab}$
and $\phi$. The Grassmann odd constraints of equation (3.14), which in
this case  are those of (4.3.10),   are the correct superspace constraints between the
vector potential and Goldstino. Thus we find that the general scheme 
given  above also works for this model. It would be interesting to
complete the calculation to the full non-linear theory.

%%%%%%%%%%%%%%%%%%%%%%%%%%%%%%%%%%%%%%%%%%%%%%%%%%%%%%%%%%%%%%%%%%%%
\medskip 
{\bf 5 $E_{11}$ formulation of brane dynamics}
\medskip
The M two  brane was constructed long ago [13] by demanding that it
possess
$\kappa$-symmetry which can be viewed as part of world volume
supersymmetry [12]. When coupled to the eleven dimensional supergravity
background the two brane also possess  local supersymmetry.  However, it
then only possess
$\kappa$-symmetry, which is essential for its consistency, if the
background supergravity fields satisfy their equations of motion [13].
Thus the two brane possess all the manifest  symmetries   of the 
background supergravity theory and   it also 
implies the   equation of motion of the background fields. This picture
is also true  for the other super branes  in ten and eleven dimensions
and in this sense they are more fundamental than the supergravity
theories to which they couple. As such, if the supergravity theories can
be extended to possess the Kac-Moody algebra
$E_{11}$ [9] one might think that this symmetry should also be present in
a suitably extended formulation of brane dynamics. Another way of arriving
at this idea is to recall that open string scattering leads to closed
strings and at low energy the latter are described by the corresponding
supergravity theories. If the supergravity theories when suitably
extended possess an
$E_{11}$ symmetry then this should  arise from  the open strings and
should be present in their dynamics. 
\par
In fact, there is a connection between brane charges and $E_{11}$.
There is considerable evidence [11,14,16]  that the brane charges 
to belong to the fundamental representation, denoted
$l_1$, of algebra
$E_{11}$ associated with the node at the end of the longest tail of the
$E_{11}$ Dynkin diagram.   The lowest level such charges are given by
[11] 
$$
P_{\underline a} ; \  Z^{\underline a\underline b}
; \  Z^{\underline a_1\ldots \underline a_5}
; \  Z^{\underline a_1\ldots \underline a_7,\underline a_8}, 
 \  Z^{\underline a_1\ldots \underline a_8}; 
\ Z^{\underline b_1 \underline b_2 \underline b_3, \underline a_1\ldots
\underline a_8};\ldots 
\eqno(5.1)$$ 
The first entry is the space-time
translations and the next two can be identified with the central charges
of the eleven dimensional supersymmetry algebra [11]. These three
quantities   are  known [17] to be the brane charges of the point
particle, two brane and five brane of M theory. The higher level objects
in the $l_1$ representation  should also  correspond to brane-like 
objects in the extended theory which possess 
$E_{11}$ symmetry. Indeed, it has been shown   that for every element 
in the $l_1$ representation  there is a
corresponding field in the adjoint representation of  $E_{11}$   
that belongs to    the
correct SL(11)  representation to allow it to be  coupled to  a   brane 
with the corresponding charge [11,16].  There has been much discussion of
the brane charges that result when the IIA string theory is dimensionally
reduced on a torus [18,19] and, in particular, it has been realised
[18,19] that the central charges that occur in the reduced supersymmetry
algebra can not form multiplets of the U-duality algebras.  However,
decomposing  the
$l_1$ representation of the algebra appropriate to dimensional reduction 
one finds [11,14] that it contains all the usual  brane charges as well 
as more exotic objects that complete the U-duality multiplets. While it is
inherent in the construction that the brane charges belong to U duality
multiplets it is encouraging that all the expected brane charges  belong
to a single 
$E_{11}$ representation and in this way one finds  an eleven dimensional
origin for the exotic charges required by U duality. 
\par
Since taking just the fields of the non-linear realisation of $E_{11}$ to
depend on just the usual coordinates of space-time, which are associated
with  $P_{\underline a}, $ breaks  $E_{11}$, it was proposed [11] that
these fields should  depend  on an infinite set of coordinates which
transform under $E_{11}$ as the  
$l_1$ representation, that is the set of coordinates 
$$
X^{\underline a} , \  X_{\underline a\underline b}
; \  X_{\underline a_1\ldots \underline a_5}
; \  X_{\underline a_1\ldots \underline a_7,\underline a_8}, 
 \  X_{\underline a_1\ldots \underline a_8}; 
\ X_{\underline b_1 \underline b_2 \underline b_3, \underline a_1\ldots
\underline a_8,};\ldots 
\eqno(5.2)$$ 
\par
From this perspective one may  suppose that the dynamics
of branes is $E_{11}$ invariant and that it is constructed from
 fields which transform in the $l_1$ representation as in  
equation (5.2).  Bosonic branes and superbranes can be viewed as 
defects in Minkowski space and superspace respectively and it is a natural
generalisation to suppose that the bosonic sector of superbranes
corresponds to a defect in a space with coordinates given in  equation
(5.2).  The dynamical fields of the brane being Goldstone bosons for the
brane charges which are spontaneously broken. 
\par
The result found earlier in this paper, namely that the world volume gauge
fields and the usual transverse scalar fields of the brane have a 
common origin in that they are Goldstone bosons for the corresponding
central charges provided encouragement for the idea of an
underlying $E_{11}$ symmetry of brane dynamics. Indeed, it implies that
these fields must occur in the group element of the non-linear
realisation of equation (3.7) in precisely the required  way to be
suitable to be  embedded in an $E_{11}$ formulation.
\par
The question of whether the hidden symmetries of the supergravity
theories also occurred in dimensional reduction of brane dynamics was
given  a partial answer in reference [20].   It was shown that a
 two brane reduced on  a four torus did have the expected  SL(5,R)
symmetry. One interesting aspect of this work was that it used a  
formulation of the two brane  which possessed an apparently adhoc 
field with two anti-symmetric indices as well as the usual  scalar
fields. However, it was noted [20] that this formulation could not
account for the expected  symmetries if the dimensional reduction was on
a torus of dimension greater than four. 
\par
Clearly,  a dimensional reduction of  the brane dynamics proposed 
in this section on a torus would possess the same symmetries as   the
eleven dimensional supergravity theory under the same reduction. 
The decomposition of the $l_1$ representation appropriate to a
dimensional  reduction   to three dimension was
given in [14] and it was shown  how the  fields of equation (5.2) 
formed multiplets of the corresponding $E_8$ symmetry group. The results
for a dimensional reduction on a smaller dimension torus can also be
readily deduced from these. For a torus of dimension four it is easy to
see that $X^i, X_{ij}$ belong to the $\underline 4$ and  
$\underline 6$ of SL(4) and so form the ${\underline 10}$ of SL(5), 
while for a torus of dimension five  the $X^i, X_{ij}, X_{i_1\ldots i_5}$
transform as the 
$\underline 5$, $\underline {10}$ and $\underline 1$ of SL(5) and so
form the $\underline {16}$ of SO(5,5). Here the indices $i,j,\dots$ are
in the directions in which the torus lies. However, for a seven torus the
coordinates corresponding to central charges of the supersymmetry algebra
will no longer suffice and one finds that 
$X^i, X_{ij}, X_{i_1\ldots i_5}, X_{i_1\ldots i_7,i},$ 
transform as  the
$\underline 7$ and $\underline {21}$,  $\underline {21} $ and $\underline
7$ of SL(7)  to form the $\underline {56}$ of $E_7$. On an eight torus one
needs coordinates up to level six to form the $\underline {248}$ of
$E_8$.  
\par
The usual formulation of the two brane involves only a 
$X^{\underline a}$ and so, at first sight,  it does not appear to 
possess an $E_{11}$ symmetry. However, we will now sketch an  $E_{11}$ 
formulation of brane dynamics and apply it to the two brane.  We will
find that at low levels it does indeed lead to the expected
dynamics. 
\par
It will prove instructive to first recall the 
 dynamics of a  bosonic p brane in a flat background is 
given by the non-linear realisation of
ISO(1,D-1)/SO(1,p)$\otimes$SO(D-p-1) [8]. 
The group element is of the form 
$$exp(X^{\underline a} P_{\underline a}) exp(\phi_a{}^{b'} J^a{}_{b'})
\eqno(5.3)$$
where the fields depend on the world volume coordinates $\xi^n$. The Cartan
form corresponding to the translations  is given by 
$\nabla_n X^{\underline a}=\partial _n X^{\underline p}\Phi_{\underline
p}{}^{\underline a}$, where $\Phi_{\underline
p}{}^{\underline a}=(e^\phi)_{\underline
p}{}^{\underline a} $. This transforms only under 
SO(1,p)$\otimes$SO(D-p-1) and as such we may set $\nabla_n X^{a'}=0$
which implies that the space-time derivatives of $X^{n'}$ are given in
terms of $\phi_a{}^{b'}$. The only covariant derivative of
$X^{\underline a}$ remaining is 
$\nabla_n X^{a}\equiv f_n{}^a$ and the action is given by $\int d^{p+1}\xi
\ \det f_n{}^a$.  Using the identity $ f_n{}^a\eta_{ab}f_n{}^b=
\nabla_n X^{\underline a}\eta_ {\underline a \underline
b}\nabla_m X^{\underline b} =\partial_n X^{\underline p}\eta _{\underline
p\underline q}\partial_m X^{\underline q}\equiv \gamma_{nm}$ we recognise
the familiar expression for the action. 
\par
We may generalise the above to include the coupling of the bosonic brane
to the gravity background [10] by taking as our group element 
$$exp(X^{\underline a} P_{\underline a}) exp(h_{\underline
a}{}^{\underline  b} K^{\underline a}{}_{\underline  b})
\eqno(5.4)$$
where $K^{\underline a}{}_{\underline  b}$ are the generators of 
GL(D) and we can identify $e_{\underline n}{}^{\underline a}=(\exp
h)_{\underline n}{}^{\underline a}$ as the vierbein from the gravity that
results from the group element [10]. The corresponding covariant
derivative of 
$X^{\underline a}$ is given by 
 $\nabla_n X^{\underline a}=\partial _n X^{\underline p} e_{\underline
p}{}^{\underline a}$. We define 
$\nabla_n X^{\underline a}\eta_ {\underline a \underline
b}\nabla_m X^{\underline b} =\partial_n X^{\underline p}g _{\underline
p\underline q}\partial_m X^{\underline q}\equiv \gamma_{nm}$. 
 The action is then given by  $\int d^{p+1}\xi \ \sqrt {-\det
\gamma_{nm}}$.  
The anti-symmetric part of 
$h_{ a}{}^{ b'}$ play the role of Goldstone boson for
the broken Lorentz symmetry. 
\par
We now sketch the $E_{11}$ non-linear realisation of  a super brane
dynamics in the supergravity background. We will consider only the
bosonic fields and work to second order in the brane coordinates
and the background supergravity fields. Hence we consider the group
element built from the algebra, denoted $E_{11}\otimes_s l_1$, which is
the semi-direct product of $E_{11}$ with generators that transform  in 
the $l_1$ representation of $E_{11}$. The group element is given by  
$$
g=exp(X^{\underline a} P_{\underline a}) exp(X_{\underline a
\underline b} Z^{\underline a \underline b})
exp(X_{\underline a_1 \ldots 
\underline a_5} Z^{\underline a_1\ldots  \underline a_5})\cdots 
exp(h_{\underline a}{}^{\underline  b} K^{\underline a}{}_{\underline 
b})
$$
$$exp({1\over 3!}
A_{\underline a_1 \underline a_2\underline a_3}
R^{\underline a_1\underline a_2\underline a_3})
exp({1\over 6!}A_{\underline a_1 \ldots 
\underline a_6} R^{\underline a_1\ldots  \underline a_6})\cdots 
$$
$$
\equiv g_1exp(h_{\underline a}{}^{\underline  b} K^{\underline
a}{}_{\underline  b})exp({1\over 3!}
A_{\underline a_1 \underline a_2\underline a_3}
R^{\underline a_1\underline a_2\underline a_3})
exp({1\over 6!}A_{\underline a_1 \ldots 
\underline a_6} R^{\underline a_1\ldots  \underline a_6})\cdots 
$$
$$\equiv g_1g_2
\eqno(5.5)$$
The fields associated with $l_1$ depend on the coordinates $\xi^n$ 
which parameterise the  brane world volume and
those of the supergravity background depend on the  fields in $l_1$. 
The commutators of these generators can be found in [9,11]. We will
in this sketch take as the local sub-algebra only the Lorentz algebra. 
\par
The Cartan forms are given by 
$$g^{-1}d g= \nabla_n X^{\underline a}P_{\underline a}
+\nabla_n X_{\underline a\underline b}Z^{\underline a\underline
b}+\nabla_n X_{\underline a\ldots \underline
a_5}Z^{\underline a\ldots \underline a_5}+ \dots+g^{-1}_2dg_2
\eqno(5.6)
$$
The latter are the  Cartan forms associated with $E_{11}$ algebra. 
We find that 
$$
\nabla_n X^{\underline a}=\partial _n X^{\underline p}
e_{\underline p}{}^{\underline a},\ 
\nabla_n  X_{\underline a\underline b}= \partial_n X_{\underline
p\underline q}(e^{-1})_{\underline a}{}^{\underline p}
(e^{-1})_{\underline b}{}^{\underline q}-
\partial _n X^{\underline p}
e_{\underline p}{}^{\underline c} A_{\underline c\underline a\underline b}
$$
$$
\nabla_n  X_{\underline a_1\ldots \underline a_5}= \partial_n
X_{\underline p_1\ldots \underline p_5}(e^{-1})_{\underline
a_1}{}^{\underline p_1} \cdots 
(e^{-1})_{\underline a_5}{}^{\underline p_5} +
\partial _n X^{\underline q}
e_{\underline q}{}^{\underline c} (A_{\underline c\underline
a_1\ldots \underline a_5}+10 A_{\underline c\underline
[a_1 \underline a_2}A_{\underline a_3\ldots \underline a_5]})
$$
$$
-20 A_{[\underline a_1 \underline a_2\underline a_3}
\partial_n  X_{\underline n_4\underline n_5}(e^{-1})_{\underline
a_4}{}^{\underline n_4}(e^{-1})_{\underline
a_5]}{}^{\underline n_5}
\eqno(5.7)$$
\par
We now apply this to the two brane taking field only to next to lowest
order. The equation which is first order in derivatives, constructed from
the Cartan form  and world volume reparmetisation invariant is given by  
$$
E_{\underline a}^n\equiv\sqrt {-\gamma} \gamma^{nm}
\nabla_n X^{\underline b}
\eta_{\underline a \underline b}
+d_1 \epsilon^{nmr}\nabla_m  X_{\underline a\underline b}
\nabla_r X^{\underline b}+\cdots =0
\eqno(5.8)$$
where
$$\gamma_{nm}\equiv \nabla_n X^{\underline a}\nabla_n 
\eta_{\underline a \underline b}X^{\underline b}+\cdots
\eqno(5.9)$$
and $d_1$ is a constant. Using the above expressions for the Cartan forms
this equation  become
$$
e _{\underline n}{}^{\underline a}E^{n}_{\underline a}=
\sqrt {-\gamma} \gamma^{nm}\partial_n X^{\underline p}
g_{\underline p\underline n}+d_1\epsilon^{nmr}
\partial_m X_{\underline n\underline p} \partial_r X^{\underline p}
-d_1 \epsilon^{nmr}\partial_m X^{\underline q}A_{\underline q \underline
n \underline p}\partial_r X^{\underline p}
+\cdots=0
\eqno(5.10)$$
Taking the derivative $\partial _n$ of this equation we find at
lowest order the correct equation of motion  for the two brane in a
constant background fields. For a non-constant background we would have
an equation which is second order in derivatives. 
We note that this equation contains an additional field $X_{\underline
n\underline p}$ which is dynamical in the sense that it comes with
derivatives, but it does not lead to more degrees of freedom as the
equation of motion is first order. Thus it is consistent with the
considerations discussed earlier in this paper. We note that the
inclusion of fermions would require more than just the Goldstino of
section four.  
\par
By wrapping the two brane in eleven dimensions on a circle one may take
the above dynamics and obtain the analogous results for the dynamcis of
the IIA string in ten dimensions. The  lowest level coordinates are
$X^{\underline a},\ a=0,1,\ldots 9$ and 
$\bar X_{\underline a}\equiv X_{\underline a 11}$. Making the
decomposition of the $l_1$ representation into $D_{10}$ representations 
by deleting the last node on the gravity line of  the Dynkin diagram one
finds [14] that the  lowest level multiplet contains the $X^{\underline
a}$ and $\bar X_{\underline a}$  which belong to the vector
representation of $D_{10}$. Hence, just keeping these fields one will find
a $D_{10}$ invariant dynamics which was discussed in reference [26]. In
this case, the additional coordinate
$\bar X_{\underline a}$ encodes the possibility of interchanging
Kaluza-Klein and winding modes. 
\par 
We now also sketch the five brane dynamics from this viewpoint. In
this case we must include the fields at the next level. The analogous
equation for the scalar fields is 
$$
 \gamma^{nm} \nabla_n X^{\underline b}
\eta_{\underline a \underline b}
+d_2 \epsilon^{nm_1\ldots m_5}\nabla_{m_1}  X_{\underline b\underline
a_1\ldots \underline a_5}
\nabla_{m_2} X^{\underline a_1}\cdots \nabla_{m_5} X^{\underline
a_4}+\cdots =0
\eqno(5.11)$$
where $d_2$ is a constant. While for the two form $X^{\underline
a\underline b}$ the equation is 
$$
\nabla_{[n} X_{\underline a\underline b} \nabla_m X^{\underline
a}\nabla_{p]} X^{\underline b}+\cdots \ {\rm \ is\  self\  dual}
\eqno(5.12)$$
Substituting for the Cartan forms one recognises the 
correct equations of motion for the five brane [25] in a constant
background up to the level considered.
Since any p brane has a coordinate in the $l_1$ representation that 
has p anti-symmetrised indices we recognise that equations (5.8) and
(5.11) are just special cases of an obvious generalisation that gives  the
equation of motion of the embedding coordinates. 
\par 
We will give a more systematic account of the low level dynamics
of branes viewed as a non-linear realisation of $E_{11}$
elsewhere. In particular, we will address the issue, glossed over here, of
what are the correct local sub-algebras which will in turn determine the
field content of the non-linear realisations. For the five brane we
expect to extend the local sub-algebra to include the Lorentz algebra and
 the automorphism discussed in equation (4.2.6) and we expect that this
will lead to the peculiar  factors of the three form field strength that
occur in the five brane dynamics. The brane dynamics in the absence of
background fields is likely to be a non-linear realisation of the $l_1$
representation together with the Cartan involution invariant sub-algebra
of $E_{11}$, or possibly a sub-algebra of this. 
\par
Despite the low level approximation of the above dynamics 
  the correct equations of motion appear naturally and, to those
with a sympathetic disposition, it may  encourage  the belief in an
underlying $E_{11}$ symmetry of brane dynamics.

%%%%%%%%%%%%%%%%%%%%%%%%%%%%%%%%%%%%%%%%%%%%%%%%%%%%%%%%%%%%%%%%%%%%%%%%
\medskip
{\bf {6. Discussion }}
\medskip
Inspired by Matrix Theory [24], it has been argued [18] that the
maximal $d+1$ dimensional 
 "super Yang-Mills theory" dimensional reduced on a  d-dimensional
torus has an $E_d, \ d\le 9$ symmetry where $E_d$ is the hidden symmetry
that appears when eleven dimensional supergravity is reduced on a
suitable dual torus of dimension $d$.
 In particular, it was found that this  hidden symmetry, or at
lest the relevant discrete part of it,    is generated in the "super
Yang-Mills theory" by  simply permuting the radii of compactification 
combined with Montonen-Olive duality. The authors of [18] raised the
question of what extension of   super Yang-Mills theory  could possess
such a symmetry. 
\par 
In this paper we have argued that   brane dynamics,
like  the supergravity theories to which it couples, can be extended to
admit an
$E_{11}$ symmetry. Indeed,  both are non-linear realisation of  $E_{11}$;
the supergravity theory being the low level approximation involving  the
adjoint representation while the usual brane dynamics is the low level
approximation from a non-linear realisation which also includes   the
$l_1$ representation. It is important to note that the hidden symmetries
are actually part of the theory before dimensional reduction and so 
appear automatically  when they are dimensionally reduced and from this
perspective it is inevitable that  the supergravity and brane dynamics,
when suitably extended, 
 have the same symmetries when dimensionally
reduced.  This is in contrast to the   mismatch between the
brane and M theory symmetries noted in [18]. 
 The approach advocated here also answers the another question raised in
[18], namely it provided a higher dimensional origin for the   exotic
charges whose  existence   in the Yang-Mills theory is implied by acting
with the analogue of the U-duality transformations on the well known
charges.  
\par
The ideas put forward in this paper are very natural from the viewpoint
of the gauge-gravity correspondences (the Maldacena conjecture) since 
the two theories could be viewed as just different faces of their common
underlying $E_{11}$ symmetry.     We would note that
 many of the checks of the Maldacena conjecture are really a
consequence of the symmetries of the gauge and gravity theories. 
The correspondence  between the supergravity and brane dynamics implied 
by their common symmetries could be exploited to give a mapping between
quantities in the two theories. It would be interesting to see if these
were the same as those that appear in  Matrix
theory [24].    
\par
The $l_1$ representation contains at level zero the 
usual space-time coordinates  $X^{\underline a}$. Using techniques
similar to those used in reference [16], it is straightforward to show,
modulo some very unexpected conspiracy discussed in that reference,  that
the
$l_1$  representation also contains elements with the same Sl(10)
representation content as the space-time coordinates at levels $n_c=11m,\
m\in {\bf Z}$. Thus it contains multiple copies of the space-time like
coordinates first found at level zero. Indeed, one can show that if a
representation of Sl(10) occurs at level $n_c$ it also occurs at level
$n_c+11$. As a result, one finds in effect multiple copies of the first
few coordinates used to describe brane dynamics. There is one  intriguing
interpretation of this repetition; it might be  the $E_{11}$ way of
encoding non-Abelian branes. One might interpreted the first set of
coordinates as belonging to a single brane and the subsequent sets as
belonging to multiple  branes. One  test of this idea would be 
to see if the multiplicities of the higher
coordinates are consistent with this picture in that  they make up an
U(m) adjoint multiplet.  While one might expect the generators $P_a$ at
level zero to commute it could well be the case that the higher level
generators with the same indices do not commute.

%%%%%%%%%%%%%%%%%%%%%%%%%%%%%%%%%%%%%%%%%%%%%%%%%%%%%%%%%%%%%%%%%%%%%%%%
\medskip 
{\bf Acknowledgments}
\par
I wish to thank Jonathan Bagger and David Olive for useful
discussions. This research was supported by a PPARC senior fellowship
PPA/Y/S/2002/001/44 and  in part by the PPARC grants  PPA/G/O/2000/00451
 and the EU Marie Curie, research training network
grant HPRN-CT-2000-00122. 
\medskip
{\bf References}
\medskip
 \item{[1]} S. Coleman, J. Wess, B. Zumino,  
{\sl Structure of Phenomenological Lagrangians. 1},
Phys.Rev. {\bf 177} (1969) 2239; 
D.  Volkov, {\sl Phenomenological Lagrangians.}
Sov.J. Particles and Nuclei {\bf 4} (1973) 3
\item{[2]} J. Hughes, J. Liu, J. Polchinski, 
Supermembranes.
Phys.Lett. {\bf 180B} (1986) 370; 
J.  Hughes, J.  Polchinski, 
{\sl Partially Broken Global Supersymmetry and the Superstring.}
Nucl.Phys. {\bf B278} (1986) 147.
\item{[3]} J. Bagger, A. Galperin, 
{\sl New Goldstone Multiplet for Partially Broken Supersymmetry.}
Phys.Rev. {\bf D55} (1997) 1091, hep-th/9608177. 
\item{[4]}
J. Bagger, A.  Galperin, 
{\sl Matter Couplings in Partially Broken Extended Supersymmetry.}
Phys.Lett. {\bf B336} (1994) 25, hep-th/9406217. 
\item{[5]} E. Ivanov and S. Krivonos, {\sl  $N=1, D=4$ supermembrane in
the coset approach}, Phys.Lett. {\bf B453} (1999) 237-, hep-th/9901003. 
\item{[6]} J. Bagger, A. Galperin, 
{\sl The Tensor Goldstone Multiplet for Partially Broken Supersymmetry.}
Phys.Lett. {\bf B412} (1997) 296.
\item {[7]} D. Volkov and V. Akulov, {\sl Is the neutrino a goldstone
particle}, Phys. Lett. {\bf 46B} (1973) 109. 
\item{[8]} P. West, {\sl Automorphisms, Non-linear Realizations and
Branes},  JHEP, 02 (2000) 024, hep-th/0001216.
\item{[9]} P. West, {\sl $E_{11}$ and M Theory}, Class. Quant.
Grav. {\bf 18 } (2001) 4443, {\tt hep-th/0104081}
\item{[10]} P.~C. West, {\sl Hidden superconformal symmetry in {M}
    theory },  JHEP {\bf 08} (2000) 007, {\tt hep-th/0005270}
\item{[11]} P. West, {\sl $E_{11}$, SL(32) and Central Charges},
Phys. Lett. {\bf B 575} (2003) 333-342, {\tt hep-th/0307098}
\item{[12]} D. Sorokin, V. Tkach and D.V. Volkov, {\it Superparticles,
	twistors and Siegel symmetry}, Mod. Phys. Lett. {\bf A4} (1989) 901;
 D. Sorokin, V. Tkach, D.V. Volkov and A. Zheltukhin, {\sl From
	superparticle Siegel supersymmetry to the spinning particle
proper-time	supersymmetry}, Phys. Lett. {\bf B259} (1989) 302.
\item{[13]} E. Bergshoeff, E. Sezgin and P.K. Townsend, 
{\sl Properties of eleven
dimensional supermembrane theory}, Ann. Phys. 185 (1988) 330; {\sl 
Supermembranes and eleven dimensional supergravity}, 
Phys. Lett. B189 (1987) 75.
\item{[14]} P. West, {\sl $E_{11}$ origin of brane charges and U-duality
  multiplets}, {\bf  JHEP} 0408 (2004) 052, hep-th/0406150.
\item {[15]} P. West and O. Baerwald, {\sl Brane Rotating Symmetries and
the Fivebrane Equations of Motion},  Phys.Lett. B476 (2000) 157-164,
hep-th/9912226. 
\item{[16]} A. Kleinschmidt and  P. West, {\sl
Representations of
${\cal G}^{+++}$ and the role of space-time}, JHEP {\bf 0402} (2004) 033,
hep-th/0312247. 
\item{[17]} J. Azcarraga, J. Gauntlett, J. Izquierdo and P.
Townsend,  {\it Topological Extensions of the Supersymmetry Algebra for
Extended Objects}, Phys. Rev. Lett. {\bf 63, no 22} (1989) 2443.  
\item{[18]} 
 S. Elitzur, A. Giveon, D. Kutasov and E.  Rabinovici,  {\it
Algebraic aspects of matrix theory on $T^d$ }, {hep-th/9707217}; 
  N. Obers,  B. Pioline and E.  Rabinovici, {\it M-theory and
U-duality on $T^d$ with gauge backgrounds}, { hep-th/9712084}; 
 N. Obers and B. Pioline,~ {\it U-duality and  
M-theory, an
algebraic approach}~, { hep-th/9812139}; 
 N. Obers and B. Pioline,~ {\it U-duality and  
M-theory}, {hep-th/9809039}. 
\item{[19]} B. de Wit and H. Nicolai, {\it Hidden symmetries,
central charges and all that}, hep-th/0011239. 
\item{[20]} M. Duff, {\sl Duality rotations in membrane theory}, 
Nucl. Phys. {\bf B347} (1990) 394. 
\item{[21]} P. S. Howe and  E. Sezgin, {\sl D=11, p=5} 
 { Phys. Lett.} B394 (1997) 62. 
\item{[22]} Gauntlett, K. Itoh and P. Townsend, {\sl Superparticle with
extrinsic curature},  { Phys. Lett.} {\bf  B238} (1990) 65. 
\item{[23]} E. Ivanov and V. Ogievetsky, {\sl The inverse Higgs
mechanism in non-linear realisations}, Teor. Mat. Fiz. {\bf 25} (1975)
164.   
\item{[24]} T. Banks, W. Fischler, S. Shenker, and L. Susskind, 
 {\sl M theory as a matrix model; a conjecture}, Phys. Rev. {\bf D55}
(1997) 5112. 
 \item{[25]} P. S. Howe, E. Sezgin, and P. C. West.
 {\sl Covariant field equations of the {M} theory five-brane.}
 { Phys. Lett.} {\bf B399} (1997) 49--59,  hep-th/9702008. 
\item{[26]} S. Cecotti, S. Ferrara and L. Girardello, Nucl. Phys. {\bf
B308} (1988) 436; M. Duff,  Nucl. Phys. {\bf B335} (1990) 610. 

\end